# Bifurcation analysis of the statics and dynamics of a logistic model with two delays

*M. Berezowski, E. Fudała*
*Silesian University of Technology*
*Institute of Mathematics*
*44-100 Gliwice, ul. Kaszubska 23, Poland,*
*e-mail address: Marek.Berezowski@polsl.pl*

**Abstract**

The mathematical-numerical analysis of a discrete dynamical model with two independent delays was performed. Such model may describe a continuous system with delays that have real rational number values. Applicable characteristic equations were derived for both a single and double cycle. The results of the analysis were illustrated by numerical examples in the form of boundary bifurcation curves and Feigenbaum's diagrams.

**Notation**

| | |
|---|---|
| $k$ | delay length |
| $K$ | feedback amplitude |
| $r$ | constant |
| $t$ | time, $s$ |
| $T$ | period, $s$ |
| $x$ | state variable |
| $\varphi$ | argument of $\lambda$ |
| $\lambda$ | eigenvalue |

*Subscripts*

| | |
|---|---|
| $0$ | delay |
| s | fixed point |

## 1. Introduction

There have been numerous publications concerning the bifurcation analysis of the solutions of dynamic models containing different types of delays [1-16]. The simplest form of a continuous model with single delay $\tau$ is the system shown in Fig.1a [2]. This system is qualitatively equivalent to the discrete model shown in Fig.1b.

The scope of this paper is the bifurcation analysis of a dynamic system with two different delays $\tau_1$ and $\tau_2$ (Fig.2a). Assuming that the ratio of the delays is a rational number, expressed as $\tau_1 / \tau_2 = m / n$, the continuous system shown in Fig.2a may be presented in a qualitatively equivalent discrete form – see Fig.2b. Such form may apply to the system of two apparatuses connected in parallel with feedback and different duration of flux flow.

The study included the mathematical and numerical analysis of the dynamics of the system shown in Fig.2b. In the first part, analytical equations determining the occurrence of various types of static and dynamic bifurcations were derived. Next, on the grounds of the characteristic equations the conditions describing both bifurcations involving stationary states as well as oscillation solutions were designated. In the second part of the study, the obtained analytical results were illustrated by numerical examples. To achieve this, the corresponding boundary bifurcation curves and steady state Feigenbaum's diagrams were formed.

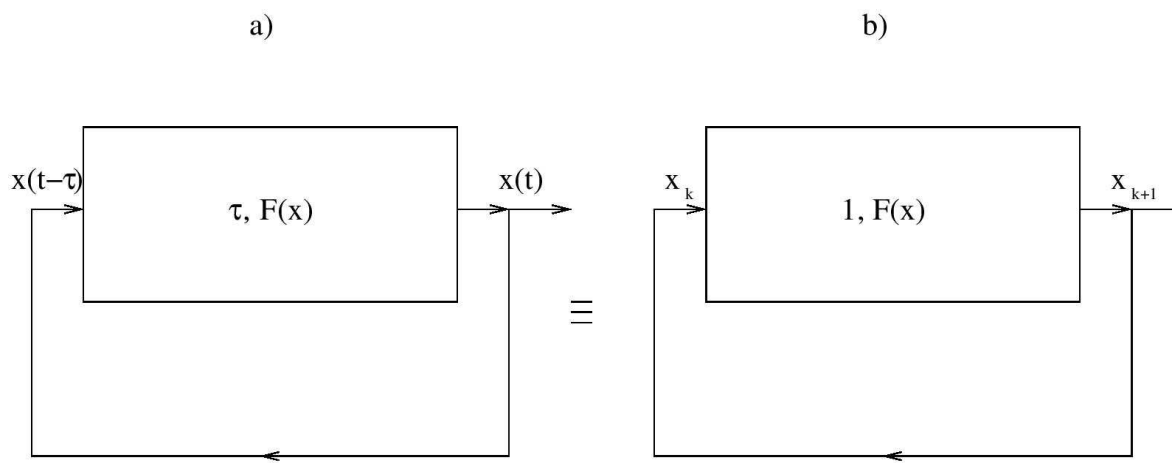

Fig.1. Schematic diagram of the system with one delay: a) continuous model, b) discrete model.

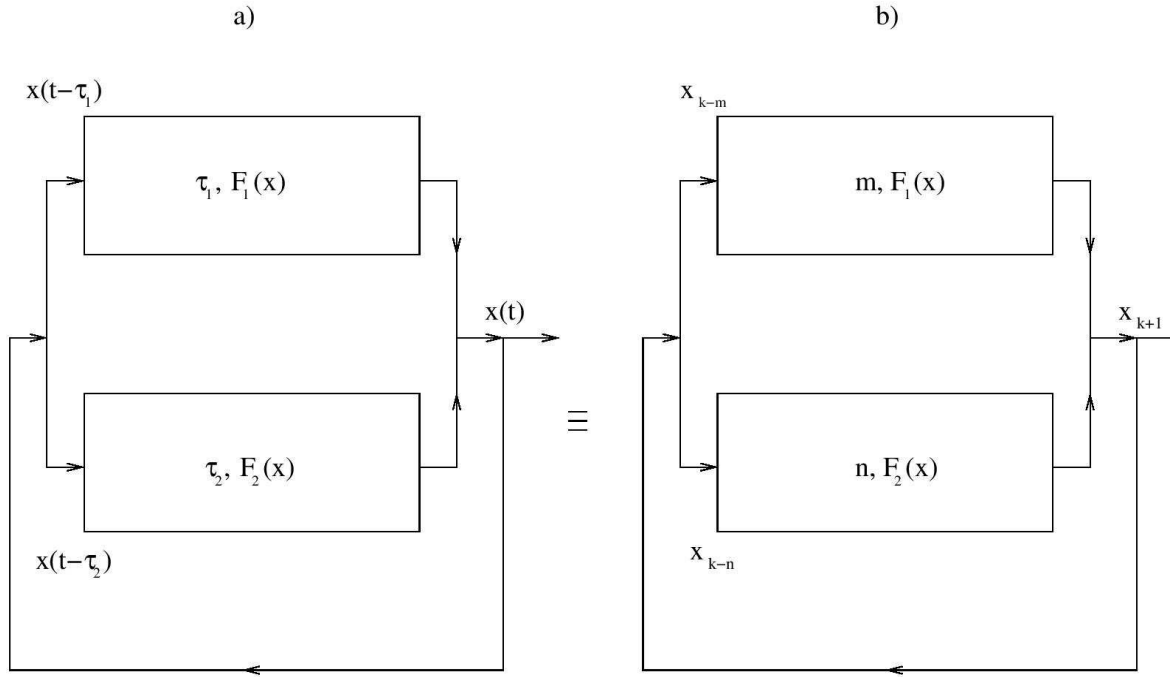

Fig.2. Schematic diagram of the system with two delays: a) continuous model, b) discrete model.

## 2. The model

The following logistics model with feedback was analyzed in [1]:

$$x(n+1) = f[x(n)] + g[x(n-k)] \tag{1a}$$

where $k$ is the delay length. It was assumed that functions $f$ and $g$ have the following form:

$$f[x(n)] = (1-K)rx(n)[1-x(n)] \tag{1b}$$

$$g[x(n-k)] = Kx(n-k) \tag{1c}$$

where $K$ is the feedback amplitude. From the physical point of view, the above model may concern a continuous system with two delays $t_0$, $(k+1)t_0$, assuming that, variable $x$ is observed in discrete moments of time $t = nt_0$. Thus, in its continuous form, equation (1a) may be expressed as:

$$x(t) = f[x(t-t_0)] + g\{x[t-(k+1)t_0]\}. \tag{2}$$

However, there are no physical obstacles for the introduction of an assumption about the ratio of the above delays not being an integer number. So, it may be assumed that the relationship between the two variables is a rational number in the following form:

$$\frac{k_2 + 1}{k_1 + 1}.$$ (3)

Accordingly, in this case, equation (2) may be written down as:

$$x(t) = f\left[x(t - t_0)\right] + g\left[x\left(t - \frac{k_2 + 1}{k_1 + 1}t_0\right)\right].$$ (4)

Defining the additional time as:

$$t_0' = \frac{t_0}{k_1 + 1}$$ (5)

equation (4) may be written down in the following form:

$$x(t) = f\left\{x\left[t - (k_1 + 1)t_0'\right]\right\} + g\left\{x\left[t - (k_2 + 1)t_0'\right]\right\}.$$ (6)

Assuming, again, that variable $x$ is observable only in discrete moments of time $t = nt_0'$ the continuous model – see (6) may be expressed in a discrete form as:

$$x(n+1) = f\left[x(n - k_1)\right] + g\left[x(n - k_2)\right].$$ (7)

Assuming, yet again, that $k_2 > k_1$, equation (7) may be formally transformed to the following system of $(k_2 + 1)$ equations:

$$
\begin{aligned}
x_1(n+1) &= f\left[x_{k_1+1}(n)\right] + g\left[x_{k_2+1}(n)\right] \\
x_2(n+1) &= x_1(n) \\
&\vdots \\
x_{k_2}(n+1) &= x_{k_2-1}(n) \\
x_{k_2+1}(n+1) &= x_{k_2}(n).
\end{aligned}
$$ (8)

By the linearization of the above system around fixed point $x_s$, the following linear approximation is obtained:

$$\bar{x}(n+1) = \bar{\bar{J}}^{(1)}(n)\bar{x}(n)$$ (9)

where $\bar{\bar{J}}^{(1)}$ is Jacobi matrix:

$$\overline{\overline{J}}^{(1)} = \begin{bmatrix} 0 & 0 & \dots & \left.\dfrac{\partial f}{\partial x_{k_1+1}}\right|_s & \dots & 0 & 0 & \left.\dfrac{\partial g}{\partial x_{k_2+1}}\right|_s \\ 1 & 0 & \dots & 0 & \dots & 0 & 0 & 0 \\ \vdots & \vdots & \ddots & \vdots & \ddots & \vdots & \vdots & \vdots \\ 0 & 0 & \dots & 0 & \dots & 1 & 0 & 0 \\ 0 & 0 & \dots & 0 & \dots & 0 & 1 & 0 \end{bmatrix}. \qquad (10)$$

In further transformations, definitions (1b) and (1c) were considered. Accordingly:

$$\left.\frac{\partial f}{\partial x_{k_1+1}}\right|_s = (1-K)r(1-2x_s) = b\,; \qquad \left.\frac{\partial g}{\partial x_{k_2+1}}\right|_s = K. \qquad (11)$$

To designate the characteristic equation of (9) it is convenient to assume that $\overline{\overline{J}}^{(1)}$ is a matrix of the following hypothetical linear differential problem:

$$\frac{du_1}{dt} = bu_{k_1+1} + Ku_{k_2+1}$$

$$\frac{d^2u_1}{dt^2} = b\frac{du_{k_1+1}}{dt} + K\frac{du_{k_2+1}}{dt} = bu_{k_1} + Ku_{k_2}$$

$$\vdots$$

$$\frac{d^{(k_1+1)}u_1}{dt^{(k_1+1)}} = b\frac{du_2}{dt} + K\frac{du_{k_2-k_1+2}}{dt} = bu_1 + Ku_{k_2-k_1+1} \qquad (12)$$

$$\frac{d^{(k_1+2)}u_1}{dt^{(k_1+2)}} = b\frac{du_1}{dt} + K\frac{du_{k_2-k_1+1}}{dt} = b\frac{du_1}{dt} + Ku_{k_2-k_1}$$

$$\vdots$$

$$\frac{d^{(k_2+1)}u_1}{dt^{(k_2+1)}} = b\frac{d^{(k_2-k_1)}u_1}{dt^{(k_2-k_1)}} + K\frac{du_2}{dt} = b\frac{d^{(k_2-k_1)}u_1}{dt^{(k_2-k_1)}} + Ku_1.$$

The above dependence makes it possible to derive the characteristic equation:

$$\lambda^{(k_2+1)} = b\lambda^{(k_2-k_1)} + K \Leftrightarrow \lambda^{(k_2+1)}\left[1 - b\lambda^{-(k_1+1)}\right] = K \qquad (13)$$

which is helpful in testing the stability of the fixed points of the logistics transformation (7).

The analysis of the stability should involve three elementary cases, connected with the evenness and unevenness of constants $k_1$ and $k_2$. Each case, as proved below, describes different relations between coefficients $r$ and $K$ determining the incidence of bifurcation. Therefore, let us assume that at Hopf bifurcation points (HB) $\left(|\lambda| = 1\right)$ constants $k_1$, $k_2$ and $k_1'$, $k_2'$ determine the following forms of the characteristic equation:

$$e^{i\varphi(k_2+1)}\left[1 - be^{-i\varphi(k_1+1)}\right] = K \qquad (14)$$

$$e^{i\varphi'(k_2'+1)}\left[1-be^{-i\varphi'(k_1'+1)}\right] = K. \tag{15}$$

Let us pose now the requirement of the above two forms being mutually equivalent, i.e. fulfilling the following conditions:

$$\varphi'(k_2'+1) = \varphi(k_2+1) \tag{16}$$

$$\varphi'(k_1'+1) = \varphi(k_1+1) \tag{17}$$

which, in turn, lead to:

$$(k_2+1)(k_1'+1) = (k_1+1)(k_2'+1). \tag{18}$$

In other words, let us pose the requirement that different values of constants $k_1$, $k_2$ and $k_1'$, $k_2'$ determine the same Hopf bifurcation point. In view of such assumption, the following three cases shall be discussed:

<u>Case A:</u> $k_1 -$ even, $k_2 -$ even

- For odd constant $k_1'$, both sides of equation (18) must be even. Accordingly, constant $k_2'$ must be an odd number.

- For even constant $k_1'$, both sides of equation (18) must be uneven. Accordingly, constant $k_2'$ must be an even number.

<u>Case B:</u> $k_1 -$ even, $k_2 -$ odd

- For odd constant $k_2'$, both sides of equation (18) must be even. Accordingly, constant $k_1'$ may be an even or an odd number. Evenness of constant $k_2'$ is impossible.

<u>Case C:</u> $k_1 -$ odd, $k_2 -$ even

- For odd constant $k_1'$, both sides of equation (18) must be even. Accordingly, constant $k_2'$ may be an odd or even number. Evenness of constant $k_1'$ is impossible.

It follows from the above analysis that the fourth case: $k_1 -$ odd, $k_2 -$ odd, is always equivalent to one of the above cases A, B or C. However, it should be noted that cases A, B, C are not mutually equivalent. To determine the values of $k_1'$ and $k_2'$, on the grounds of given $k_1$ and $k_2$, the following formula (arising from) (18) may be applied:

$$k_1' = N\frac{k_1+1}{GCD}-1, \qquad k_2' = N\frac{k_2+1}{GCD}-1 \tag{19}$$

where $GCD$ is the greatest common divisor of numbers $\left(k_1+1\right)$ and $\left(k_2+1\right)$, whereas $N$ is any natural number.

Considering (19) and assuming the equivalence of solutions, the following relation is derived:

$$\lambda^{(k_j+1)} = (\lambda')^{(k_j'+1)} = (\lambda')^{\frac{N}{GCD}(k_j+1)}; \qquad j=1,2 \qquad (20)$$

which indicates, that if $\lambda$ is the eigenvalue of a given solution, $(\lambda')^{\frac{N}{GCD}}$ is the eigenvalue of the equivalent solution. Thus:

$$\lambda = (\lambda')^{\frac{N}{GCD}} \Rightarrow |\lambda|e^{i\varphi} = |\lambda'|^{\frac{N}{GCD}} e^{i\varphi'\frac{N}{GCD}} \Rightarrow \begin{cases} |\lambda| = |\lambda'|^{\frac{N}{GCD}} \\ \varphi = \dfrac{N}{GCD}\varphi'. \end{cases} \qquad (21)$$

It follows from (21) that if the oscillation period of a given solution is $T$, the oscillation period of the equivalent solution is:

$$T' = \frac{N}{GCD}T \qquad (22)$$

since both solutions differ only in terms of the time axis. Hence, for example, if $k_1 = 2$, $k_2 = 5$ assuming that $N = 2$, on the grounds of equations (19) $k_1' = 1$ and $k_2' = 3$ are derived. This means that the discrete system (7) with delays equal to 2 and 5 and oscillation period $T$, is equivalent to the discrete system (7) with delays equal to 1 and 3 and oscillation period $T' = \dfrac{2}{3}T$.

Apart from HB bifurcation, two special cases of bifurcation may occur in a discrete dynamic system: static bifurcation (LP) $(\lambda = +1)$ and flip bifurcation (FB) $(\lambda = -1)$ – which leads to abrupt (irregular) oscillations [2]. It is easy to prove that the discussed logistics model (7) offers two different fixed points:

$$x_{s1} = 0, \quad x_{s2} = \frac{r-1}{r}. \qquad (23)$$

The analysis presented below shall concern point $x_{s2}$ and its neighborhood. It follows from characteristic equation (13) that static bifurcation (LP) of this point occurs for $r = 1$, irrespective of the value of parameter $K$ and autonomously from case A, B or C. Dynamic bifurcation FB occurs for different values of $r$ and $K$, depending on the above mentioned cases A, B, C. Accordingly,

in case A, bifurcation FB occurs for

$$r = \frac{3-K}{1-K} \qquad (24)$$

in B, for

$$r = 3 \qquad\qquad\qquad (25)$$

whereas in case C, for

$$r = \frac{1 - 3K}{1 - K}. \qquad\qquad\qquad (26)$$

Each of the above mentioned bifurcations FB causes the loss of stability of fixed point $x_{s2}$, leading to the generation of successive fixed points $x_{s3}$ and $x_{s4}$ which are the amplitudes of a double cycle.

The logistics transformation of the double cycle may, in turn, be expressed by the following system of recurrent equations:

$$x(n+1) = f\big[x(n-k_1)\big] + g\big[x(n-k_2)\big] \qquad\qquad (27)$$

$$x(n+2) = f\big[x(n+1-k_1)\big] + g\big[x(n+1-k_2)\big]. \qquad\qquad (28)$$

In the designation of the fixed points of transformation (27) – (28) the above mentioned three cases should be discussed:

<u>Case A:</u> ( $k_1 -$ even, $k_2 -$ even)

$$x(n+1) = x(n-1) = ... = x(n-k_1+1) = ... = x(n-k_2+1) = x_{s3} \qquad (29)$$

$$x(n+2) = x(n) = ... = x(n-k_1) = ... = x(n-k_2) = x_{s4}. \qquad (30)$$

Thus, considering that (27) – (28):

$$x_{s3} = f(x_{s4}) + g(x_{s4}) \qquad\qquad\qquad (31)$$

$$x_{s4} = f(x_{s3}) + g(x_{s3}). \qquad\qquad\qquad (32)$$

Including (1b) and (1c) in the above equations, the fixed points are expressed as:

$$x_{s3} = \frac{(1-K)r + K + 1 - \sqrt{[(1-K)r + K + 1]\,[(1-K)r + K - 3]}}{2(1-K)r} \qquad (33)$$

$$x_{s4} = \frac{(1-K)r + K + 1 + \sqrt{[(1-K)r + K + 1]\,[(1-K)r + K - 3]}}{2(1-K)r}. \qquad (34)$$

<u>Case B:</u> ( $k_1 -$ even, $k_2 -$ odd)

$$x(n+1) = x(n-1) = ... = x(n-k_1+1) = ... = x(n-k_2) = x_{s3} \qquad (35)$$

$$x(n+2) = x(n) = ... = x(n-k_1) = ... = x(n-k_2+1) = x_{s4}. \qquad (36)$$

Thus, after considering (27) – (28):

$$x_{s3} = f(x_{s4}) + g(x_{s3}) \tag{37}$$

$$x_{s4} = f(x_{s3}) + g(x_{s4}). \tag{38}$$

Including (1b) and (1c) in the above equations, the fixed points are expressed as:

$$x_{s3} = \frac{r + 1 - \sqrt{(r+1)(r-3)}}{2r} \tag{39}$$

$$x_{s4} = \frac{r + 1 + \sqrt{(r+1)(r-3)}}{2r} \tag{40}$$

<u>Case C:</u> ($k_1$ – odd, $k_2$ – even)

$$x(n+1) = x(n-1) = ... = x(n-k_1) = ... = x(n-k_2+1) = x_{s3} \tag{41}$$

$$x(n+2) = x(n) = ... = x(n-k_1+1) = ... = x(n-k_2) = x_{s4}. \tag{42}$$

Thus, accounting that (27) – (28):

$$x_{s3} = f(x_{s3}) + g(x_{s4}) \tag{43}$$

$$x_{s4} = f(x_{s4}) + g(x_{s3}). \tag{44}$$

Including (1b) and (1c) in the above equations, the fixed points are expressed as:

$$x_{s3} = \frac{(1-K)r - 1 - K - \sqrt{[1 + K - (1-K)r]\,[1 - 3K - (1-K)r]}}{2(1-K)r} \tag{45}$$

$$x_{s4} = \frac{(1-K)r - 1 - K + \sqrt{[1 + K - (1-K)r]\,[1 - 3K - (1-K)r]}}{2(1-K)r}. \tag{46}$$

The derived formulas, determining the fixed points of transformation (27) – (28) make up the so called: branches of periodic solutions of a double cycle on the steady-states diagram. Certainly, bifurcation points type HB and FB may also appear on the branches. Their position may be designated from another characteristic equation, concerning, this time, problem (27) – (28). Assuming that $k_2 > k_1$, Jacobi matrix of the problem has the following form:

$$\bar{\bar{J}}^{(2)} = \begin{bmatrix} 0 & \dots & \left.\frac{\partial f}{\partial x_{k_1}}\right|_{s3} & 0 & \dots & 0 & \left.\frac{\partial g}{\partial x_{k_2}}\right|_{s3} & 0 \\ 0 & \dots & 0 & \left.\frac{\partial f}{\partial x_{k_1+1}}\right|_{s4} & \dots & 0 & 0 & \left.\frac{\partial g}{\partial x_{k_2+1}}\right|_{s4} \\ 1 & \dots & 0 & 0 & \dots & 0 & 0 & 0 \\ \vdots & \ddots & \vdots & \vdots & \ddots & \vdots & \vdots & \vdots \\ 0 & \dots & 0 & 0 & \dots & 1 & 0 & 0 \end{bmatrix}. \quad (47)$$

Likewise, to determine the characteristic equation of (27) – (28), it is convenient to assume that $\bar{\bar{J}}^{(2)}$ is a matrix of the following hypothetical linear differential system:

$$\frac{du_1}{dt} = b_1 u_{k_1} + K u_{k_2}$$

$$\frac{du_2}{dt} = b_2 u_{k_1+1} + K u_{k_2+1}$$

$$\frac{du_3}{dt} = u_1 \qquad\qquad (48)$$

$$\vdots$$

$$\frac{du_{k_2+1}}{dt} = u_{k_2-1}$$

where:

$$b_1 = \left.\frac{\partial f}{\partial x_{k_1}}\right|_{s3} = (1-K)r(1-2x_{s3})$$

$$b_2 = \left.\frac{\partial f}{\partial x_{k_1+1}}\right|_{s4} = (1-K)r(1-2x_{s4}) \qquad (49)$$

$$\left.\frac{\partial g}{\partial x_{k_2}}\right|_{s3} = \left.\frac{\partial g}{\partial x_{k_2+1}}\right|_{s4} = K.$$

Depending on case A, B or C, the system of equations (48) is transformed as follows:

**A:**

$$\frac{d^{\left(\frac{k_2}{2}\right)}u_1}{dt^{\left(\frac{k_2}{2}\right)}} = b_1\frac{d^{\left(\frac{k_2-k_1}{2}\right)}u_2}{dt^{\left(\frac{k_2-k_1}{2}\right)}} + Ku_2 \qquad\qquad (50)$$

$$\frac{d^{\left(\frac{k_2}{2}+1\right)}u_2}{dt^{\left(\frac{k_2}{2}+1\right)}} = b_2\frac{d^{\left(\frac{k_2-k_1}{2}\right)}u_1}{dt^{\left(\frac{k_2-k_1}{2}\right)}} + Ku_1 \qquad\qquad (51)$$

-the characteristic equation of which is:

$$b_1 b_2 \lambda^{(k_2 - k_1)} + (b_1 + b_2) K \lambda^{\left(\frac{k_2 - k_1}{2}\right)} + K^2 = \lambda^{(k_2 + 1)}. \qquad (52)$$

If $\dfrac{k_2 - k_1}{2}$ is an even number, it follows from equation (52) that bifurcation FB$^{(2)}$ occurs for:

$$r = \frac{1 - K + \sqrt{6}}{1 - K}. \qquad (52a)$$

If $\dfrac{k_2 - k_1}{2}$ is an odd number, it follows from equation (52) that bifurcation FB$^{(2)}$ occurs for:

$$r = \frac{1 - K + \sqrt{6 + 4K(1 + K)}}{2}. \qquad (52b)$$

**B:**

$$\frac{d^{\left(\frac{k_2 + 1}{2}\right)} u_1}{dt^{\left(\frac{k_2 + 1}{2}\right)}} = b_1 \frac{d^{\left(\frac{k_2 + 1 - k_1}{2}\right)} u_2}{dt^{\left(\frac{k_2 + 1 - k_1}{2}\right)}} + K u_1 \qquad (53)$$

$$\frac{d^{\left(\frac{k_2 + 1}{2}\right)} u_2}{dt^{\left(\frac{k_2 + 1}{2}\right)}} = b_2 \frac{d^{\left(\frac{k_2 - 1 - k_1}{2}\right)} u_1}{dt^{\left(\frac{k_2 - 1 - k_1}{2}\right)}} + K u_2 \qquad (54)$$

-the characteristic equation of which is:

$$\left( \lambda^{\left(\frac{k_2 + 1}{2}\right)} - K \right)^2 = b_1 b_2 \lambda^{(k_2 - k_1)}. \qquad (55)$$

If $\dfrac{k_2 + 1}{2}$ is an even number, it follows from equation (55) that bifurcation FB$^{(2)}$ occurs for:

$$r = 1 + \sqrt{6}. \qquad (55a)$$

If $\dfrac{k_2 + 1}{2}$ is an odd number, it follows from equation (55) that bifurcation FB$^{(2)}$ occurs for:

$$r = \frac{1 - K + \sqrt{5(1 - K)^2 + (1 + K)^2}}{1 - K}. \qquad (55b)$$

**C:**

$$\frac{d^{\left(\frac{k_2}{2}\right)} u_1}{dt^{\left(\frac{k_2}{2}\right)}} = b_1 \frac{d^{\left(\frac{k_2 - 1 - k_1}{2}\right)} u_1}{dt^{\left(\frac{k_2 - 1 - k_1}{2}\right)}} + K u_2 \qquad (56)$$

$$\frac{d^{\left(\frac{k_2}{2}+1\right)}u_2}{dt^{\left(\frac{k_2}{2}+1\right)}} = b_2\frac{d^{\left(\frac{k_2+1-k_1}{2}\right)}u_2}{dt^{\left(\frac{k_2+1-k_1}{2}\right)}} + Ku_1 \qquad (57)$$

-the characteristic equation of which is:

$$b_1 b_2 \lambda^{(k_2-k_1)} + \lambda^{(k_2+1)} - (b_1+b_2)\lambda^{\left(\frac{2k_2-k_1+1}{2}\right)} = K^2. \qquad (58)$$

If $\frac{k_1-1}{2}$ is an even number, it follows from equation (58) that bifurcation FB$^{(2)}$ occurs for:

$$r = \frac{1-K-\sqrt{2(3K^2+2K+2)}}{1-K}. \qquad (58a)$$

If $\frac{k_1-1}{2}$ is an odd number, it follows from equation (58) that bifurcation FB$^{(2)}$ occurs for:

$$r = \frac{1-K(1+\sqrt{6})}{1-K}. \qquad (58b)$$

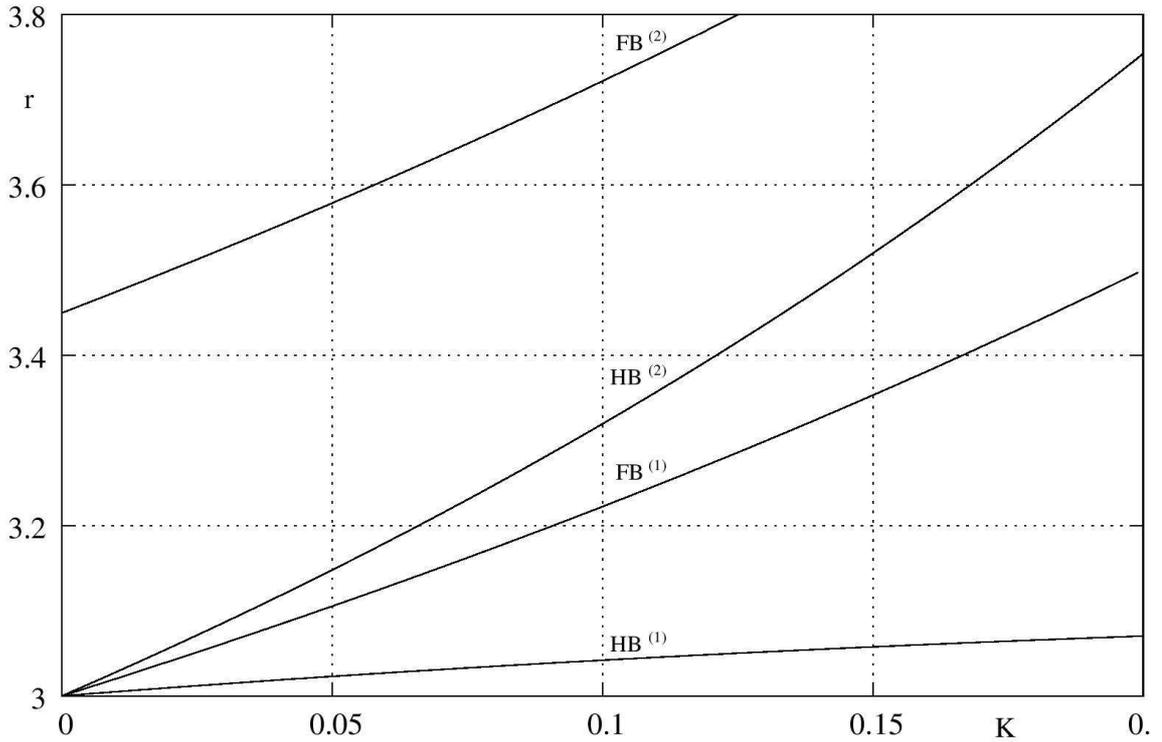

Fig.3. Boundary bifurcation lines: $k_1 = 2$, $k_2 = 6$.

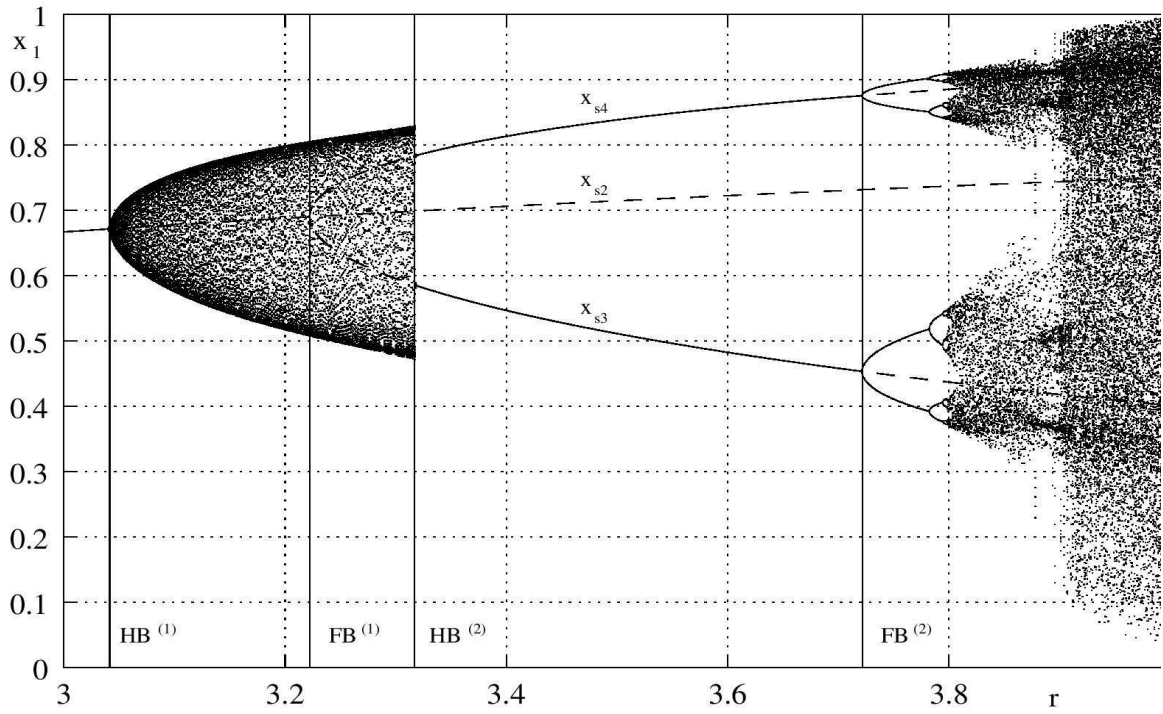

Fig.4. Diagram of steady states: $k_1 = 2$, $k_2 = 6$, $K = 0.1$.

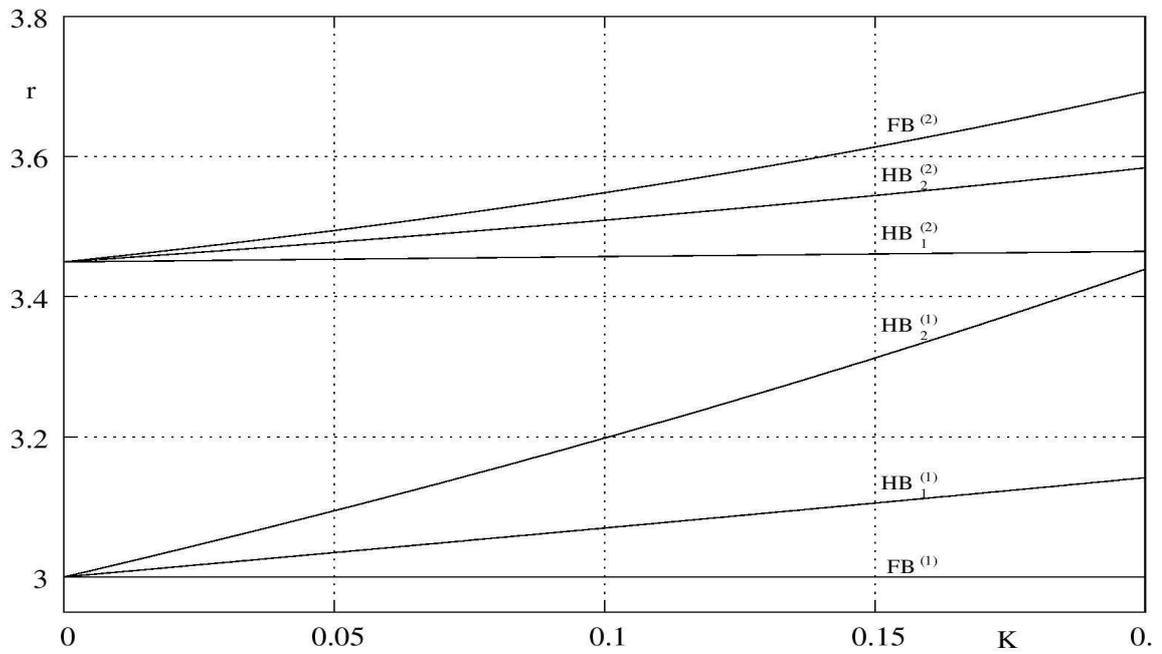

Fig.5. Boundary bifurcation lines: $k_1 = 4$, $k_2 = 5$.

## 3. Exemplary calculations

This Chapter contains some selected examples of calculations illustrating the analytical results discussed in the previous Chapter. They refer to cases: A, B and C, respectively.

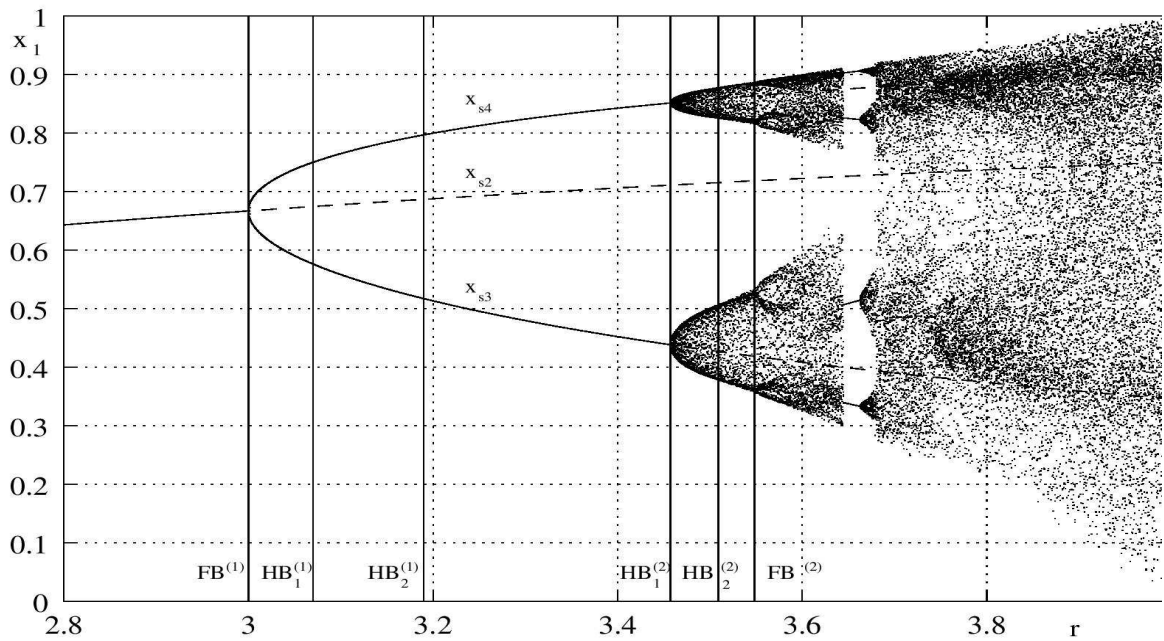

Fig.6. Diagram of steady states: $k_1 = 4$, $k_2 = 5$, $K = 0.1$.

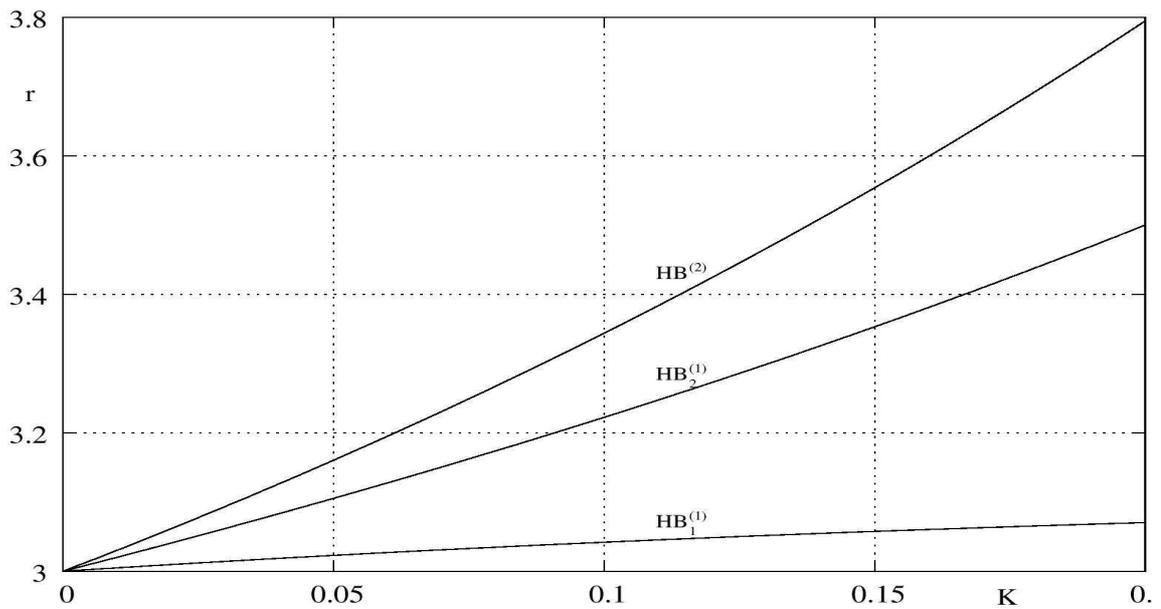

Fig.7. Boundary bifurcation lines: $k_1 = 3$, $k_2 = 6$.

Case A:

Assuming that: $k_1 = 2$, $k_2 = 6$, Fig.3 shows the bifurcation lines corresponding to this case. Thus, line HB$^{(1)}$ designated from equation (14) is a set of single cycle Hopf bifurcation points (with φ as a parameter). Line FB$^{(1)}$ designated from equation (24) a set of single cycle flip bifurcation points. Line HB$^{(2)}$ designated from equation (52) $\left( \left| \lambda \right| = 1 \right)$ is a set of double cycle Hopf bifurcation points (with $\varphi$ as a parameter). Line FB$^{(2)}$ designated from equation (52a) is a set of double cycle flip bifurcation points.

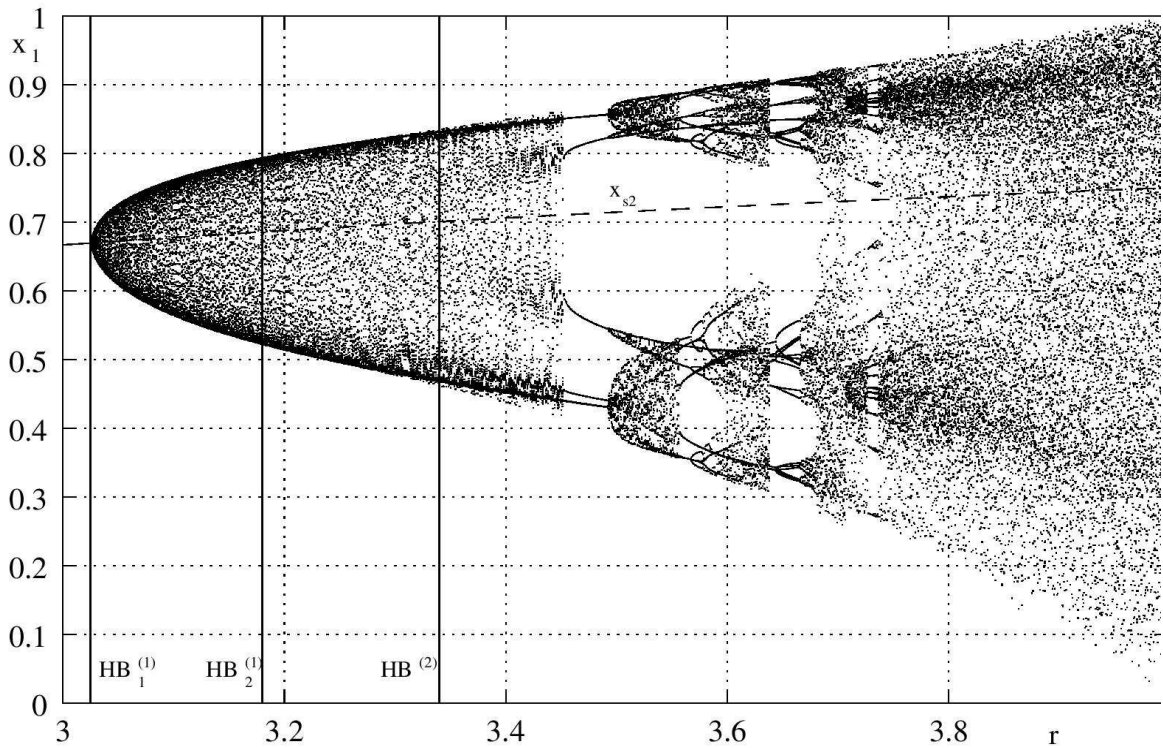

Fig.8 . Diagram of steady states: $k_1 = 3$, $k_2 = 6$, $K = 0.1$.

Exemplary Feigenbaum's bifurcation diagram for $K = 0.1$ is shown in Fig.4. The line of single cycle fixed points $x_{s2}$ was designated from equation (23), whereas the line of double cycle fixed points $x_{s3}$ and $x_{s4}$ was designated from equations (33) and (34). The vertical lines mark particular bifurcation points, corresponding to the values from Fig.3. As may be observed, in the range of HB$^{(1)} < r <$ HB$^{(2)}$ the discrete system generates quasiperiodic oscillations, whereas in the range of $\text{HB}^{(2)} < r < \text{FB}^{(2)}$ the oscillations are oneperiodic. They are initiated by bifurcation $\text{FB}^{(1)}$. In the range of $\text{FB}^{(1)} < r < \text{HB}^{(2)}$ the oscillations are unstable. For high values of parameter $r$ the system generates chaotic oscillations. Stable states are marked with continuous lines, whereas unstable states by broken lines.

Case B:

Assuming that: $k_1 = 4$, $k_2 = 5$, Fig.5 shows the bifurcation lines corresponding to this case. Thus, line $\text{FB}^{(1)}$ satisfying equation (25) is a set of single cycle flip bifurcation points. Lines $\text{HB}_1^{(1)}$ and $\text{HB}_2^{(1)}$ designated from equation (14) are a set of single cycle Hopf bifurcation points (with $\varphi$ as a parameter). Lines $\text{HB}_1^{(2)}$ and $\text{HB}_2^{(2)}$ designated from equation (55) $\left( |\lambda| = 1 \right)$ are set of double cycle Hopf bifurcation points (with $\varphi$ as a parameter). Line $\text{FB}^{(2)}$ designated from equation (55b) is a set of double cycle flip bifurcation points. Exemplary bifurcation diagram for $K = 0.1$ is shown in Fig.6. The line of single cycle fixed points $x_{s2}$ was designated from equation (23), whereas the line of double cycle fixed points $x_{s3}$ and $x_{s4}$ was designated from equations (39) and (40). The vertical lines mark particular bifurcation points, corresponding to the values from Fig.5. As may be observed in this case, bifurcations $\text{HB}_1^{(1)}$, $\text{HB}_2^{(1)}$, $\text{HB}_2^{(2)}$ and $\text{FB}^{(2)}$ have no impact on the generation of stable oscillations in the system. However, this does not mean that it is impossible to select the initial values of the state variables to ensure the significance of these bifurcations.

Case C:

Assuming that: $k_1 = 3$, $k_2 = 6$, Fig.7 shows the bifurcation lines corresponding to this case. Thus, lines $\text{HB}_1^{(1)}$ and $\text{HB}_2^{(1)}$ designated from equation (14) are a set of single cycle Hopf bifurcation points (with $\varphi$ as a parameter). Line $\text{HB}^{(2)}$ designated from equation (58) $\left( |\lambda| = 1 \right)$ is a set of double cycle Hopf bifurcation points (with $\varphi$ as a parameter). Exemplary bifurcation diagram for $K = 0.1$ is shown in Fig.8. The line of single cycle fixed points $x_{s2}$ was designated from equation (23). The vertical lines mark particular bifurcation points, corresponding to the values from Fig.7. As may be observed in this case, bifurcations $\text{HB}_2^{(1)}$ and $\text{HB}^{(2)}$ have no impact on the generation of stable oscillations in the system. However, as in the case above, the initial values of the state variables may be selected in such a way as to ensure the significance of these bifurcations.

## 4. Concluding remarks

The scope of the study is the mathematical and numerical analysis of a logistics model with two independent delays, expressed by means of natural numbers. Such model offers a qualitative mathematical description of a continuous system with two different delays, expressed by means of real rational numbers. The conducted analysis involved mainly bifurcation phenomena that might occur in the tested system. Thus, applicable characteristic equations were derived for a single and double cycle. The derived equations determine Hopf bifurcation generated by fixed stationary points as well as by fixed points of periodic solutions. Both evenness and unevenness of particular delays, as well as evenness and unevenness of the expressions noted by means of the delays, were proved to have a significant impact on the quality of the solution. The analytical results were illustrated by numerical examples in the form of applicable bifurcation curves and Feigenbaum's diagrams.